\documentclass[a4paper,twocolumn,showpacs,prb]{revtex4}%

\usepackage{amsfonts}
\usepackage{amsmath}
\usepackage{amssymb}
\usepackage{graphicx}%
\providecommand{\U}[1]{\protect \rule{.1in}{.1in}}

\begin{document}
\title{Enhanced visibility of graphene: effect of one-dimensional photonic crystal}

\author{Kai Chang, J. T. Liu and J. B. Xia}
\affiliation{NLSM, Institute of Semiconductors, Chinese Academy of
Sciences, P. O. Box 912, Beijing 100083, China}
\author{N. Dai}
\affiliation{National Laboratory for Infrared Physics, Shanghai
Institute of Technical Physics, Chinese Academy of Sciences,
Shanghai 200083, China}


\begin{abstract}
We investigate theoretically the light reflectance of a graphene
layer prepared on the top of one-dimensional Si/SiO$_{2}$ photonic
crystal (1DPC). It is shown that the visibility of the graphene
layers is enhanced greatly when 1DPC is added, and the visibility
can be tuned by changing the incident angle and light wavelengths.
This phenomenon is caused by the absorption of the graphene layer
and the enhanced reflectance of the 1DPC.

\end{abstract}
\maketitle

Graphene consists of a two-dimensional honeycomb lattice of carbon
atoms and has been attracting attention recently due to its
remarkable electronic properties and its potential application in
nanoelectronics \cite{Novoselov}. Graphene exhibits high crystal
quality, an exotic Dirac-type spectrum, and ballistic transport on a
submicro scale. Graphene samples are usually fabricated by a
micromechnical cleavage of graphite. It is difficult to distinguish
the single graphene layer from many graphitic pieces, even utilizing
the atomic force, scanning-tunneling, and electron microscopes. A
recent experiment demonstrated that the graphene visibility depends
on both the thickness of the SiO$_{2}$ layer and the light
wavelength\cite{Geim}. They found that specific thicknesses (300nm
and 100nm) are most suitable for its visual detection for the normal
light incidence and attribute this phenomenon to the opacity of the
graphene layer. Although the relative difference of the reflectance
[the contrast $C$ in Ref. (2)] is enhanced significantly, the
absolute difference of the light reflectance is still quite low
because it is determined by the weak absorption of the graphene
layer. In order to enhance the visibility of graphene, i.e., the
absolute and relative difference of the light reflectance of the
graphene layer, we propose to prepare the graphene layer on the top
of \textit{Si/SiO} $_{2}$ one-dimensional photonic crystal(1DPC).
This 1DPC shows a high dielectric contrast at the
\textit{Si/SiO}$_{2}$ interface ($\Delta n\approx2.3$) producing a
high reflectance at normal incidence, and can be fabricated by
different techniques, e.g., the separation-byimplanted-oxygen
technique\cite{Namavar}, sputtering\cite{Sputting} combined with
solid-source Si molecular beam epitaxy\cite{Ishikawa}, and
plasma-enhanced chemical vapor deposition\cite{Lipson}.

In this Letter, we investigate theoretically the light reflectance
of a graphene layer prepared on the top of \textit{Si/SiO}$_{2}$
1DPC, as shown
schematically in Fig. 1(a). We consider an asymmetric 1DPC: \textit{A}$_{0}%
$\textit{(AB)}$_{l}$, where $l$ is an integer denoting the \textit{l}-th
layer. All layers are nonmagnetic ($\mu=1$) and are characterized by their
permittivities $\varepsilon_{A}$(\textit{SiO}$_{2}$ layer), $\varepsilon_{B}%
$(\textit{Si} layer), and their thicknesses satisfy $\sqrt{\varepsilon_{A}%
}d_{A}=\sqrt{\varepsilon_{B}}d_{B}=\lambda/4$ where $\lambda$ is the
wavelength required by the observation. The thickness of the top
\textit{SiO}$_{2}$ layer is $d=\lambda/2\sqrt{\varepsilon_{A}}$. We
find that the difference between the reflectance of the graphene
layers with 1DPC can be enhanced greatly, even one order of
magnitude larger than that without 1DPC. Furthermore, the visibility
of the graphene can be tuned by the incident angle.

\begin{figure}[ptb]\includegraphics[width=0.9\columnwidth,clip]{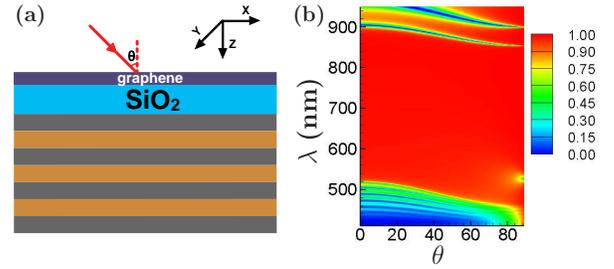}
\caption{(Color online) (a) Schematic of the graphene layer prepared
on the top of one-dimensional
SiO$_{2}$(SiO$_{2}/\operatorname{Si}$)$^{10}$ photonic crystal; (b)
photonic bandgap of SiO$_{2}$(SiO$_{2}/\operatorname{Si}$)$^{10}$
1DPC for different
incident angles.}%
\end{figure}
We consider a light shedding on the graphene layer prepared on the
top of Si/SiO$_{2}$ 1DPC with a incident angle $\theta$ from air
(refractive index, \textit{n}$_{0}$ = 1). Based on the Maxwell
equations for a monochrome light
propagating in the medium, we have%
\begin{equation}
\left \{
\begin{array}
[c]{c}%
\mathbf{k}\cdot \mathbf{D}=\mathbf{k}\cdot \varepsilon \varepsilon_{0}%
\mathbf{E}=0,\\
\mathbf{k}\cdot \mathbf{B}=\mathbf{k}\cdot \mu \mu_{0}\mathbf{H}=0,\\
\mathbf{k}\times \mathbf{E}=\omega \mathbf{B=}\omega \mu \mu_{0}\mathbf{H,}\\
\mathbf{k}\times \mathbf{H}=-\omega \mathbf{D=-}\omega \varepsilon \varepsilon
_{0}\mathbf{E,}%
\end{array}
\right.
\end{equation}
where
$\varepsilon=\varepsilon_{r}+i\varepsilon_{i}(\varepsilon_{0})$ is
the permittivity of the material (vacuum), $\mu(\mu_{0})$ the
magnetic permeability of material (vacuum), and $\omega$ the angular
frequency of the incident light.

For the TE polarization, the electric field is in the \textit{x} direction,
$\mathbf{E}_{l}\mathbf{=E}_{l}\mathbf{(}y\mathbf{,}z\mathbf{)e}_{\mathbf{x}},$
and the magnetic field is in the \textit{y-z} plane, $\mathbf{H}%
=H_{y}(y,z)\mathbf{e}_{\mathbf{x}}+H_{z}(y,z)\mathbf{e}_{z}$, where
$\mathbf{e}_{i}(i=x,y,z)$ are the unit vectors in the \textit{x}, \textit{y},
and \textit{z} directions, respectively. The reflected and transmitted
electric fields from the 1DPC\ are, respectively.%
\begin{align}
\mathbf{E}_{0}\mathbf{(}y\mathbf{,}z\mathbf{)}  &  =[A_{0}e^{ik_{z}z}%
+B_{0}e^{-ik_{z}z}]e^{-ik_{y}y}\mathbf{e}_{x},\\
\mathbf{E}_{N+1}\mathbf{(}y\mathbf{,}z\mathbf{)}  &  =A_{N+1}e^{ik_{z}%
z-ik_{y}y}\mathbf{e}_{x}.
\end{align}

\begin{figure}[ptb]
\includegraphics[width=0.9\columnwidth,clip]{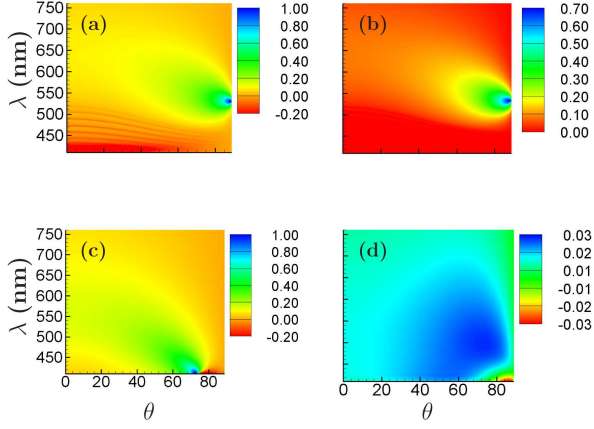}
\caption{(Color online) The contour plots of the relative and
absolute contrast $C_{r}$ [(a) and (c)] and $C_{a}$ [(b) and (d)] of
single graphene layer as a function of the wavelength $\lambda$ and
the incident angle $\theta$ with [(a) and (b)] and without [(c) and
(d)] the
1DPC.}%
\end{figure}
The electric fields of the monochrome light beam in the
\textit{l}-th layer is given by
\begin{align}
\mathbf{E}_{l}\mathbf{(}y\mathbf{,}z\mathbf{)}  &  =[A_{l}e^{ik_{z}z}%
+B_{l}e^{-ik_{z}z}]e^{-ik_{y}y}\mathbf{e}_{x},\\
H_{l}\mathbf{(}y\mathbf{,}z\mathbf{)}  &  =\frac{1}{\omega \mu_{l}\mu_{0}%
}\mathbf{k}_{l}\times \mathbf{E}_{l},\\
&  =\frac{1}{\omega \mu_{l}\mu_{0}}(k_{z}\mathbf{e}_{y}-k_{y}\mathbf{e}%
_{z})[A_{l}e^{ik_{z}z}+B_{l}e^{-ik_{z}z}]e^{-ik_{y}y},
\end{align}
where $k_{z}=\sqrt{k^{2}-k_{y}^{2}}$ in the medium. The wavevector
$k=\omega/c$ in a vacuum ($c$ is the speed of light in a vacuum),
but is generally complex in a medium. The electric fields of the
light in the \textit{l}-th layer are related to the incident fields
by the transfer matrix
utilizing the boundary condition $\mathbf{n}\times(\mathbf{H}_{1}%
-\mathbf{H}_{2})=0$, $\mathbf{n}\times(\mathbf{E}_{1}-\mathbf{E}_{2})=0$,%
\begin{equation}
\left(
\begin{array}
[c]{c}%
A_{l}\\
B_{l}%
\end{array}
\right)  =\left(
\begin{array}
[c]{cc}%
T_{11} & T_{12}\\
T_{21} & T_{22}%
\end{array}
\right)  \left(
\begin{array}
[c]{c}%
A_{0}\\
B_{0}%
\end{array}
\right)  ,
\end{equation}

The reflectance $r$ is defined as $r=\frac{\varepsilon_{0}}{\varepsilon_{N+1}%
}|\frac{T_{22}}{T_{21}}|^{2}$, and $N$ is the total layer number of
the 1DPC ($N=10$ in our calculation). The absolute and relative
contrasts describing the difference between the reflectance with and
without the graphene layer are
defined as%

\begin{align}
C_{a}  &  \equiv r(0)-r(n),\\
C_{r}  &  \equiv \lbrack r(0)-r(n)]/r(0),
\end{align}
where $r(n)$ denotes the reflectance of the sample with
\textit{n}-layer graphene. The latter ($C_{r}$) is the same as the
definition of the contrast $C$ in Ref. [2]. In order to observe the
graphene layer experimentally, both the relative contrast $C_{r}$
and the difference in reflectance between the structures with and
without the graphene layer, i.e., the absolute contrast $C_{a}$,
should be large.

The parameters used in our calculation are: the refraction index $n_{g}%
\approx2.6-1.3i$, and the permittivity
$\varepsilon_{g}=n_{g}^{2}=5.07-6.76i$ for graphene layer
\cite{Geim,Palik}, the real and imaginary parts of the permittivity
$\varepsilon_{A}$ and $\varepsilon_{B}$ for \textit{Si} and
\textit{SiO}$_{2}$ depending on the wavelength\cite{Sipara}, the
thicknesses of the \textit{SiO}$_{2}$ and \textit{Si} layers are
$d_{A}=\lambda/4n_{A}$ ($n_{A}=1.46$ for $SiO_{2}$ at
$\lambda=650nm$) and $d_{B}=\lambda/4n_{B}$ ($n_{B}=3.77$ for $Si$
$\lambda=650nm$)\cite{Sipara}, respectively.

\begin{figure}[ptb]
\includegraphics[width=0.975\columnwidth,clip]{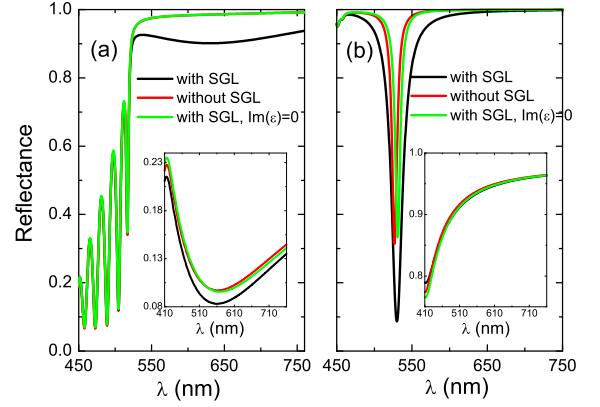}
\caption{(Color online) The reflectance as function of the light
wavelength for different structures at
normal ((a)) and oblique ((b),$\theta=89^o$) incidences. The insets show the reflectance of structures without the 1DPC}%
\end{figure}

Fig. 1(b) depicts the photonic band gap of 1DPC\ without the
graphene layer for different incident angles $\theta$. The photonic
band gap of
\textit{SiO}$_{2}$(\textit{SiO}$_{2}$/\textit{Si})$^{10}$ 1DPC
increases as the incident angle $\theta$ increases. The decrease of
the reflectance in the band gap at very large incident angle is
caused by the absorption in the \textit{Si} layers. Fig. 2 shows the
contour plots of the relative and absolute contrast $C_{r}$ and
$C_{a}$ of single graphene layer (SGL) with and without 1DPC as a
function of the light wavelength $\lambda$ and the incident angle
$\theta$. In this figure we find that that the difference between
the reflectances with and without the 1DPC exhibits a maximum at
specific light wavelengths and large incident angles $\theta$. This
light wavelength is in the band gap of the 1DPC, i.e., the high
reflection region [see Fig. 1(b)]. The maxima of the contrasts
$C_{r}$ and $C_{a}$ come both from the absorption or opacity of the
graphene layer and the maximum reflection of the eigenmode of the
1DPC, i.e., $\lambda=650nm$ at normal incidence. This figure
demonstrates that the reflection of the light is enhanced greatly
compared to that without the 1DPC, consequently leading to a large
difference in the absolute contrast $C_{a}$ between the two samples,
i.e., $C_{a}$\ with the 1DPC is one order magnitude (actually 20
times) larger than that without the 1DPC [see Fig. 2 (c) and (d)].
This enhancement should be helpful for the observation of the
graphene. In addition, the contrasts also increase significantly
with increasing the incident angle at a specific wavelength, and the
maxima of the contrast shift to the shorter wavelengths such as
$\lambda=525nm$ at larger incident angles. This is due to the
enhancement of the absorption of the graphene layer and the increase
of the optical path length at larger incident angles $\theta$. The
light wavelength corresponding to the maxima of the contrasts can be
tuned by changing the layer thicknesses of the 1DPC and the incident
angle $\theta$. This also provides us a new way to observe the
graphene in the light frequency region.

\begin{figure}[ptb]
\includegraphics[width=0.9\columnwidth,clip]{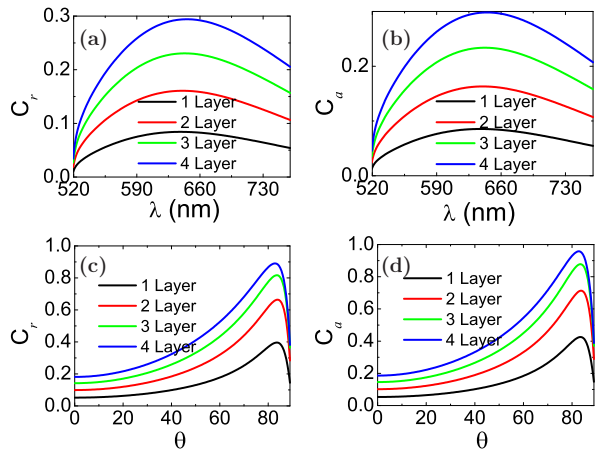}
\caption{(Color online) The relative [(a) and (c)] and absolute
contrasts [(b) and (d)] as a function of
the light wavelength and the incident angle for different graphene layers.}%
\end{figure}

In order to understand the big difference of the absolute contrast
$C_a$ between the samples with and without the 1DPC (see Fig. 2(b)
and 2(d)), we calculated the reflectance of the different
structures, i.e., the $SiO_{2}$ layer prepared on Si substrate with
and without SGL, and the $SiO_{2}$ layer prepared on the 1DPC with
and without SGL. In Fig. 3 (a) we find that the reflectances of the
1DPC with and without SGL are both enhanced greatly compared to that
without the 1DPC (see the inset of Fig. 3(a)) since the light
wavelength locates at the band gap of the 1DPC. If the absorption of
the SGL is neglected, i.e., $Im(\varepsilon)=0$ (see the green lines
in Fig. 3(a)), the reflectance of the 1DPC with SGL is almost same
as that of the 1DPC without the SGL, i.e., the very small absolute
contrast $C_a$. Fig. 3(a) demonstrates clearly that the absorption
of SGL and the enhanced reflection of light by the 1DPC are both
important for the large absolute contrast $C_{a}$ (see Fig. 2(b)).
The light reflectance of the system is determined not only by the
1DPC, but also the absorption of the graphene layer. At the large
incident angle case (see Fig. 3(b)), the absolute contrast
$C_{a}=r(0)-r(n)$ can be enhanced significantly compared with that
at the normal incidence nearby the valley ($\lambda\approx510nm$) of
the refelctance of the 1DPC which is caused by the absorption of the
Si and $SiO_2$ layers in the 1DPC, but becomes negligible small at
other light wavelengths.

Considering the multi-layer graphene prepared on the top of the
1DPC, we plot the absolute and relative contrasts as a function of
the wavelength and the incident angle (see Fig. 4). The multi-layer
graphene is modeled by the corresponding number of planes separated
by $d_{1}=0.34nm$ (the thickness of the single graphene layer). From
this figure one can see that the contrasts $C_{r}$ and $C_{a}$
exhibit significant differences among the graphene layers with
different thicknesses and the maximum as a function of the light
wavelength and the incident angle. The difference increases as the
number of the graphene layers increases. This feature makes it
possible to distinguish the number of the graphene layers.

In summary, we demonstrate theoretically that the visibility of the
graphene layers prepared on the top of
SiO$_{2}$(SiO$_{2}/\operatorname{Si}$)$^{10}$ 1DPC can be enhanced
greatly, especially at the large incident angles and specific
wavelengths in the photonic band gap. The large differences in the
reflectance make it possible for the graphene layers of different
thicknesses to be more easily observed and distinguished
experimentally.
\begin{acknowledgments}
This work was supported by the NSFC Grant Nos. 60525405, 90301007 and 10334030.
\end{acknowledgments}

\newpage

\end{document}